\documentclass[fleqn, useAMS, usenatbib]{mn2e}

\usepackage{times}
\usepackage{epsfig, amssymb}
\usepackage{graphics}
\usepackage{listings}
\usepackage{graphicx}
\usepackage{natbib}
\usepackage{eufrak}
\usepackage{mathrsfs}
\usepackage{eucal}
\usepackage{amsmath}
\usepackage{amssymb}
\usepackage{fontenc}
\usepackage{aas_macros}%
\usepackage{soul}

\def\etal{et al.~} 

\def\galform{{\texttt{GALFORM}\ }}

\newcommand{\ergsec}	{\ifmmode \,\mathrm{erg~s}^{-1} \else erg~s$^{-1}$\fi}
\newcommand{\yr}	{\ifmmode \mathrm{yr} \else yr\fi}
\newcommand{\mpc}	{\ifmmode \,\mathrm{Mpc}^{-3} \else \,Mpc$^{-3}$\fi}
\newcommand{\Msun}	{\ifmmode \,\mathrm M_{\odot} \else $\,\mathrm M_{\odot}$\fi}
\newcommand{\Mbh}	{\ifmmode M_{\mathrm{BH}} \else $M_{\mathrm{BH}}$\fi}
\newcommand{\Mseed}	{\ifmmode M_{\mathrm{BH,seed}} \else $M_{\mathrm{BH,seed}}$\fi}
\newcommand{\Mbulge}	{\ifmmode M_{\mathrm{Bulge}} \else $M_{\mathrm{Bulge}}$\fi}
\newcommand{\Mhalo}	{\ifmmode M_{\mathrm{Halo}} \else $M_{\mathrm{Halo}}$\fi}
\newcommand{\Mhaloeff}	{\ifmmode M_{\mathrm{Halo,eff}} \else $M_{\mathrm{Halo}}$\fi}
\newcommand{\Medd}	{\ifmmode \dot{M}_{\mathrm{Edd}} \else $\dot{M}_{\mathrm{Edd}}$\fi}
\newcommand{\Lbol}	{\ifmmode L_{\mathrm{bol}} \else $L_{\mathrm{bol}}$\fi}
\newcommand{\Ledd}	{\ifmmode L_{\mathrm{Edd}} \else $L_{\mathrm{Edd}}$\fi}
\newcommand{\Lcool}	{\ifmmode L_{\mathrm{cool}} \else $L_{\mathrm{cool}}$\fi}
\newcommand{\Lsun}	{\ifmmode L_{\odot} \else $L_{\odot}$\fi}
\newcommand{\Lsx}	{\ifmmode L_{\mathrm{SX}} \else $L_{\mathrm{SX}}$\fi}
\newcommand{\Mbj}	{\ifmmode M_{b_{\rm J}} \else $M_{b_{\rm J}}$\fi}
\newcommand{\bj}		{\ifmmode b_{\rm J} \else $b_{\rm J}$\fi}
\newcommand{\Lxray}	{\ifmmode L_{\mathrm{xray}} \else $L_{\mathrm{HX}}$\fi}
\newcommand{\ledd}	{\ifmmode \lambda_{\mathrm{Edd}} \else $\lambda_{\mathrm{Edd}}~$\fi}
\newcommand{\lgledd}	{\ifmmode \log_{10}\lambda_{\mathrm{Edd}} \else $\log_{10}\lambda_{\mathrm{Edd}}$\fi}
\newcommand{\rhobh}	{\ifmmode \rho_{\mathrm{BH}} \else $\rho_{\mathrm{BH}}$\fi}
\newcommand{\mdot}	{\ifmmode \dot{m} \else $\dot{m}$\fi}
\newcommand{\fhalo}	{\ifmmode f_{\mathrm{Halo}}^{\mathrm{act}} \else $f_{\mathrm{Halo}}^{\mathrm{act}}$\fi}
\newcommand{\fvis}	{\ifmmode f_{\mathrm{vis}} \else $f_{\mathrm{vis}}$\fi}
\newcommand{\fobsc}	{\ifmmode f_{\mathrm{obsc}} \else $f_{\mathrm{obsc}}$\fi}
\newcommand{\fq}		{\ifmmode f_{\mathrm{q}} \else $f_{\mathrm{q}}$\fi}
\newcommand{\beff}		{\ifmmode b_{\mathrm{eff}} \else $b_{\mathrm{eff}}$\fi}
\newcommand{\fbh}	{\ifmmode F_{\mathrm{BH}} \else $F_{\mathrm{BH}}$\fi}
\newcommand{\LxMhplane}	{\ifmmode \log_{10}L_{\mathrm{HX}}-\log_{10}M_{\mathrm{Halo}} \else $\log_{10}L_{\mathrm{HX}}-\log_{10}M_{\mathrm{Halo}}$\fi}

\title[AGN fuelling modes and clustering]{Constraints on black hole fuelling modes from the clustering of X-ray AGN} 
\author[Fanidakis \etal]{
N.~Fanidakis$^{1}$\thanks{E-mail: fanidakis@mpia.de},
A.~Georgakakis$^{2,3}$,  
G.~Mountrichas$^{3}$,  
M.~Krumpe$^{4,5}$,  
C.~M.~Baugh$^{6}$,  
\newauthor 
C.~G.~Lacey$^{6}$,  
C.~S.~Frenk$^{6}$,  
T.~Miyaji$^{4,7}$,  
A.~J.~Benson$^{8}$ \\
$^1$ Max-Planck-Institut f\"ur Astronomie, K\"onigstuhl 17, D-69117 Heidelberg, Germany.\\
$^2$ Max Planck Institut f\"ur Extraterrestrische Physik, Giessenbachstraße, 85748 Garching, Germany.\\
$^3$ National Observatory of Athens, V. Paulou \& I. Metaxa, 11532, Greece.\\
$^4$ University of California, San Diego, Center for Astrophysics and
Space Sciences, 9500 Gilman Drive, La Jolla, CA 92093-0424, USA.\\
$^5$ European Southern Observatory, ESO Headquarters,
Karl-Schwarzschild-Stra\ss e 2, 85748 Garching bei M\"unchen, Germany.\\
$^6$ Institute for Computational Cosmology, Department of Physics, University of Durham, Science Laboratories, South Road, Durham, DH1 3LE, UK.\\
$^7$ Instituto de Astronom\'ia, Universidad Nacional Aut\'onoma de Mexico,
103km Carret. Tijunana-Ensenada, Ensenada, 22860, Mexico.\\
$^8$ Carnegie Observatories, 813 Santa Barbara Street, Pasadena, CA 91101, U.S.A.
}

\begin{document} 

\maketitle
\label{firstpage}

\begin{abstract} 
We present a clustering analysis of X-ray selected AGN by compiling X-ray samples from the literature and re-estimating the dark matter (DM) halo masses of AGN in a uniform manner. We find that moderate luminosity AGN ($L_{\rm 2-10\,keV}\simeq10^{42}-10^{44}\ergsec$) in the $z\simeq0-1.3$ Universe are typically found in DM haloes with masses of $\sim10^{13}\Msun$. We then compare our findings to the theoretical predictions of the coupled galaxy and black hole formation model \texttt{GALFORM}. We find good agreement when our calculation includes the hot-halo mode of accretion onto the central black hole. This type of accretion, which is additional to the common cold accretion during disk instabilities and galaxy mergers, is tightly coupled to the AGN feedback in the model. The hot-halo mode becomes prominent in DM haloes with masses greater than $\sim10^{12.5}\Msun$, where AGN feedback typically operates, giving rise to a distinct class of moderate luminosity AGN that inhabit rich clusters and superclusters. Cold gas fuelling of the black hole cannot produce the observationally inferred DM halo masses of X-ray AGN. Switching off  AGN feedback in the model results in a large population of luminous quasars ($L_{\rm 2-10\,keV} > 10^{44}\ergsec$) in DM haloes with masses up to $\sim10^{14}\Msun$, which is inconsistent with the observed clustering of quasars. The abundance of hot-halo AGN decreases significantly in the $z\simeq3-4$ universe. At such high redshifts, the cold accretion mode is solely responsible for shaping the environment of moderate luminosity AGN. Our analysis supports two accretion modes (cold and hot) for the fuelling of supermassive black holes and strongly underlines the importance of AGN feedback in cosmological models both of galaxy formation and black hole growth.
\end{abstract}

\section{Introduction}

In the last decade numerous lines of evidence have combined to suggest that Active Galactic Nuclei (AGN) play an important, although not well understood role, in the formation and evolution of galaxies \citep{Alexander2012}. Therefore, understanding the conditions under which supermassive black holes (BHs) grow their mass across cosmic time is important not only for placing the accretion history of the Universe in a physical context but also for completing our picture of galaxy evolution. Open questions include the nature of the fuelling of supermassive black holes (BHs), the triggering mechanisms of AGN activity and the impact of the energy output of the central engine on galaxy scales.

Observationally, one approach used to address these points is via population studies of AGN as a function of cosmic time and accretion
luminosity. In particular, properties such as the morphology, star-formation history, stellar mass distribution, and large-scale environment of the galaxies that host AGN hold important clues about the physical processes that dominate the growth of BHs at different epochs (\citealt{Hopkins2009}; \citealt{Georgakakis2011}). However, the intense luminosity of AGN can easily outshine their host galaxies, rendering the study of the host's properties challenging and prone to systematics. Despite efforts to mitigate this problem, e.g., by improving analysis techniques (e.g., \citealt{Jahnke2004}; \citealt{Jahnke2007}) or by observing at wavebands where the underlying galaxy dominates \citep[e.g., far-infrared,][]{Santini2012}, contamination by AGN radiation remains a serious source of bias in studies of the hosts of active BHs. One of the few observables that is immune to this effect is the clustering of AGN, which can be interpreted in terms of their distribution in dark-matter (DM) haloes.

The interpretation of the observed properties of AGN to gain insight into the physical processes at play requires comparison with models for the cosmological evolution of AGN. These include numerical simulations (e.g., \citealt{Sijacki2007}; \citealt{Khalatyan2008}; \citealt{DiMatteo2008}; \citealt{Booth2009}), semi-empirical methods (e.g, \citealt{Hopkins2008}, and references within) or semi-analytical models (SAMs, \citealt{Malbon2007}; \citealt{Marulli2008}; \citealt{Fontanot2011}; \citealt{Fanidakis2011}). The latter combine N-body simulations of the hierarchical clustering of DM with the analytical descriptions of key physical processes in the baryons, such as gas cooling/heating, star-formation and accretion onto BHs. The advantage of this approach is its computational ease, which allows predictions to be made for the populations of AGN and galaxies, for different input parameters and adopted physical processes (e.g., BH accretion trigger, galaxy/AGN interplay). The semi-analytic approach is therefore well suited for understanding the conditions under which BH grow their mass, and the impact this has on the host galaxy.

SAMs which postulate that BHs grow in major galaxy merger events predict AGN DM halo masses of up to few times $10^{12}\,h^{-1}\Msun$, almost independent of redshift and accretion luminosity \citep{Bonoli2009}. This is in good agreement with clustering measurements of powerful, UV bright QSOs (e.g., \citealt{Croom2005}; \citealt{daAngela2008}; \citealt{Ross2009}) and underlines the importance of cold gas accretion during mergers, at least for a subset of the AGN population. At the same time, however, and contrary to merger model predictions, X-ray selected AGN, which dominate the accretion history of the Universe, are generally found in more massive DM haloes \citep[$\Mhalo\approx 10^{12.5}-10^{13.5} h^{-1}\Msun$,][]{Cappelluti2012}. This suggests that major mergers cannot be the only channel for building BHs and that alternative fuelling modes are likely to be in operation, perhaps even dominating AGN activity at certain cosmic epochs and accretion luminosities (e.g., \citealt{Allevato2011}; \citealt{Mountrichas2013}).

One SAM which includes multiple modes for growing BHs is \galform \citep{Cole2000}. Originally developed to study the cosmological evolution of galaxies, \galform has been extended recently to model  AGN activity and feedback (\citealt{Bower2006}; \citealt{Fanidakis2011}; \citealt{Fanidakis2012}). BHs grow during the different stages of the evolution of their hosts by accreting either cold gas during starbursts (dominated by disk instabilities in these models) or diffuse hot gas from a quasi-hydrostatic halo. The two fuelling modes build up the mass and spin of the BH, and the resulting accretion power regulates the gas cooling and subsequent star formation in the galaxy. This model can reproduce the observed relation between the mass of the BH and the mass of the galaxy bulge, the radio luminosity function of radio-loud AGN \citep{Fanidakis2011} as well as the luminosity function of the overall AGN population in different bands (optical, X-ray, bolometric) over a wide range of redshifts  \citep[$0\lesssim z\lesssim 6$,][]{Fanidakis2012}.

This paper extends the comparisons between observations and the \galform model predictions to the DM halo masses of X-ray AGN at different redshifts and accretion luminosities. Throughout the paper we adopt $\Omega_{\mathrm{m}}=0.227$, $\Omega_{\mathrm{b}}=0.045$, $\Omega_{\mathrm{\Lambda}}=0.728$, $\rm H_{0}=70\,km\,s^{-1}\,Mpc^{-1}$ and $\sigma_8=0.81$. The paper is organised as follows. In Section~2 we describe our method for calculating the host DM halo masses of the X-ray selected AGN in our observational samples. In Section~3 we explore the BH fuelling modes and their accretion properties in \galform and present the resulting X-ray luminosity--DM halo mass correlation. In Section~4 we compare the \galform predictions for the DM halo masses of X-ray AGN with the observations. In Section~5 we discuss the main points of our analysis. Finally, the paper is concluded in Section~6.

\section{Observational determination of X-ray selected AGN dark-matter halo masses}

Compiling a set of homogeneously estimated host DM halo masses for X-ray selected AGN from the literature to compare with the SAM predictions is challenging. Diverse methods have been employed by different groups to measure AGN clustering including, for example, the angular auto-correlation function \citep[e.g.,][]{Basilakos2005}, the real space auto-correlation function \citep[e.g.,][]{Gilli2009} and the cross-correlation with galaxies \citep[e.g.,][]{Coil2009}. Different approaches are also adopted to
infer the bias and DM halo mass from the clustering signal. Halo Occupation Distribution (HOD, e.g., \citealt{Miyaji2011}; \citealt{Krumpe2012}) modeling is a powerful way to infer clustering information from observational data. However, in the case of small sample sizes and noisy data, such as those available in many X-ray AGN studies, this method does not provide any significant advantage. As a result, many groups choose to use less physically motivated single power-law fits to describe the clustering signal of AGN and infer their bias and DM halo masses.

In this paper we use only observational studies that infer the clustering of X-ray AGN using either the real-space auto-correlation function (\citealt{Cappelluti2010}; \citealt{Gilli2009}; \citealt{Starikova2011}) or their real-space cross-correlation function with galaxies (\citealt{Coil2009}; \citealt{Krumpe2010b}; \citealt{Krumpe2012}; \citealt{Allevato2011};  \citealt{Mountrichas_Georgakakis2012};  \citealt{Mountrichas2013}). We also exclude from the analysis DM halo mass measurements inferred from wide redshift intervals, e.g. $z\approx0-3$. In studies where HOD modelling is adopted to analyse the clustering signal (\citealt{Starikova2011}; \citealt{Krumpe2012}) we use the inferred DM halo masses directly and simply scale them to $\rm
H_{0}=70\,km\,s^{-1}\,Mpc^{-1}$. In studies that fit power laws to the clustering signal, we re-estimate the bias in a uniform manner using the relation
\begin{equation} 
b_{\rm{AGN}}=\frac{\sigma_{8,\rm{AGN}}}{\sigma_8(z)},
\label{eqn:bs8}
\end{equation}
\noindent where $\sigma_{8,\rm{AGN}}$, $\sigma_8(z)$ are the rms fluctuations of the X-ray AGN and DM density distribution respectively, within a sphere of comoving radius $8\,h^{-1}\,\rm{Mpc}$. $\sigma_{8,\rm{AGN}}$ is determined from the clustering length $r_0$ and power-law exponent $\gamma$ of the AGN real-space auto-correlation function as 
\begin{equation}
\sigma^{2}_{\rm{AGN}}=J_{2}(\gamma)\left(\frac{r_0}{8h^{-1}\,\rm{Mpc}}\right)^{\gamma},
\end{equation}
where $J_{2}$ is an integral over the correlation function, which, for a power law, simplifies to 
\begin{equation}
J_{2}(\gamma)=\frac{72}{(3-\gamma)(4-\gamma)(6-\gamma)2^{\gamma}}.
\end{equation}
The values of $\gamma$ and $r_0$ are taken from the relevant publication for each sample. The error on the bias is determined from the uncertainty in the clustering length and power-law exponent. We then infer the DM halo mass from the AGN bias assuming the ellipsoidal collapse model of \cite{Sheth2001}, as described by \cite{daAngela2008} and \cite{vandenBosch2002}. The DM halo mass estimated in this way is an effective halo mass, since it represents an average over the distribution of halo masses for each AGN sample.

Table \ref{table:xray_samples} presents the DM halo mass, mean redshift, and average $2-10\,\rm{keV}$ X-ray luminosity for each AGN sample used to compare against the predictions of the \galform model. We note that the samples in Table \ref{table:xray_samples} are selected in different X-ray energy bands. The X-ray luminosities of each sample are converted to the $2-10\,\rm{keV}$ band assuming an intrinsic power-law X-ray spectrum with photon index of $\Gamma=1.9$ \citep{Nandra1994}. Also, the \cite{Krumpe2012} AGN sample includes powerful sources selected in the ROSAT $0.1-2.4\,\rm{keV}$ band. The observed flux in that band has a large soft-excess contribution that is not representative of the underlying intrinsic power-law X-ray spectrum. For these objects we use the template X-ray spectrum of powerful radio-quiet QSOs from \cite{Krumpe2010a} to account for the soft-excess contribution and extrapolate the observed flux in the ROSAT band to the intrinsic power-law luminosity in the $2-10\,\rm{keV}$ energy range.

\begin{table*}
\caption{DM halo mass measurements for galaxies that host X-ray AGN taken from the literature. Columns are: (1) The median redshift of the sample; (2) the redshift range of the sample; (3) the logarithmic value of the derived DM halo mass; (4) the average $2-10\,\rm{keV}$ X-ray luminosity of the AGN sample (the errors represent the range of luminosities in each sample); (5) the methodology used to determine the clustering signal, i.e., power-law fit (PL) or halo occupation distribution (HOD); (6) the name of the X-ray sample; (7) the reference to the relevant clustering paper for each sample.}
\label{table:xray_samples}
\begin{center}
\scalebox{1.0}{
\begin{tabular}{ccccccc}
	\hline 
	z & z range & $\log\Mhalo$ & $\log\Lxray$ & Methodology & Sample & Reference $^{a}$ \\	 & & [$h^{-1}\Msun$] & [erg s$^{-1}]$ & & & \\
	 \hline 
	 \hline 
	 \\
	0.10 & 0.03-0.20 & 13.16$^{+0.18}_{-0.23}$ & 42.1$^{+1.2}_{-0.8}$ & cross/PL & XMM/SDSS & Mountrichas \& Georgakakis (2012) \\[5pt]
	0.69 & 0.40-0.90 & 13.83$^{+0.18}_{-0.26}$ & 42.5$^{+1.5}_{-1.4}$ & cross/PL & AEGIS/COSMOS/ECDFS & Mountrichas et al. (2013) \\[5pt]
	0.97 & 0.70-1.40 & 13.06$^{+0.22}_{-0.31}$ & 42.9$^{+1.7}_{-2.0}$ & cross/PL & AEGIS/COSMOS/ECDFS & Mountrichas et al. (2013) \\[5pt]
	0.13 & 0.07-0.16 & 13.37$^{+0.15}_{-0.16}$ & 42.8$^{+0.6}_{-0.5}$ & cross/HOD & RASS/SDSS & Krumpe et al. (2012)$^{**}$ \\[5pt]
	0.27 & 0.16-0.36 & 13.32$^{+0.15}_{-0.14}$ & 43.4$^{+0.6}_{-0.4}$ & cross/HOD & RASS/SDSS & Krumpe et al. (2012)$^{**}$ \\[5pt]
	0.42 & 0.36-0.50 & 12.66$^{+0.38}_{-0.33}$ & 43.8$^{+0.5}_{-0.3}$ & cross/HOD & RASS/SDSS & Krumpe et al. (2012)$^{**}$ \\[5pt]
	0.80 & - & 13.27$^{+0.06}_{-0.06}$ & 43.53$^{+0}_{-0}$ & cross/HOD & XMM/COSMOS & Allevato et al. (2011)$^{*}$ \\[5pt]
	0.90 & 0.70-1.40 & 13.14$^{+0.18}_{-0.22}$ & 43.2$^{+1.5}_{-1.2}$ & cross/PL & AEGIS & Coil et al. (2009) \\[5pt]
 	0.05 & 0.00-0.15 & 13.20$^{+0.13}_{-0.24}$ & 43.5$^{+1.5}_{-2.5}$ & auto/PL & BAT & Cappelluti et al. (2010) \\[5pt]
	0.94 & 0.40-1.60 & 12.95$^{+0.20}_{-0.35}$ & 43.4$^{+1.7}_{-1.5}$ & auto/PL & COSMOS & Gilli et al. (2009) \\[5pt]
	0.37 & 0.17-0.55 & 12.72$^{+0.12}_{-0.15}$ & 42.7$^{+0.8}_{-0.7}$ & auto/HOD & Bo\"otes & Starikova et al. (2011)$^{*}$ \\[5pt]
	0.74 & 0.55-1.00 & 13.08$^{+0.11}_{-0.15}$ & 43.4$^{+0.7}_{-0.6}$ & auto/HOD & Bo\"otis & Starikova et al. (2011)$^{*}$ \\[5pt]
	1.28 & 1.00-1.63 & 12.85$^{+0.19}_{-0.35}$ & 44.0$^{+0.4}_{-0.5}$ & auto/HOD & Bo\"otis & Starikova et al. (2011)$^{*}$ \\[5pt]
	1.30 & - & 13.22$^{+0.08}_{-0.08}$ & 43.53$^{+0}_{-0}$ & cross/HOD & XMM/COSMOS & Allevato et al. (2011)$^{*}$ \\[5pt]

	 \hline	 	 	 	 
\end{tabular}}
\end{center}
 {Notes. $^{a}$References with one and two asterisks indicate samples in the
$0.5-2\,\rm{keV}$ and $0.1-2.4\,\rm{keV}$ soft X-ray bands respectively. Conversion to the hard band is
 performed by assuming an intrinsic
power-law X-ray spectrum with photon index of $\Gamma=1.9$
\citep{Nandra1994}.} 
\end{table*}

\section{The \galform model}

\texttt{GALFORM} calculates galaxy properties using differential equations to model the processes that describe the large and small scale physics involved in galaxy formation and BH growth. Among the most prominent are (i) the formation and evolution of DM haloes in the $\Lambda$ cold DM cosmology ($\Lambda$CDM), (ii) gas cooling and disk formation in DM haloes, (iii) star formation, supernova feedback and chemical enrichment in galaxies, (iv) accretion onto BHs and AGN feedback, and (v) the formation of bulges during galactic disk instabilities and galaxy mergers. The model has been successful in reproducing many observations including the luminosity and stellar mass function of galaxies \citep{Bower2006}, the number counts of submillimeter galaxies \citep{Baugh2005}, the evolution of Lyman-break galaxies (\citealt{Lacey2011}; \citealt{Gonzalez-Perez2013}), the clustering of Ly-$\alpha$ emitters \citep{Orsi2008}, the HI and CO mass functions (\citealt{Kim2011}; \citealt{Lagos2011}; \citealt{Lagos2012}), the space density of radio-loud AGN \citep{Fanidakis2011} and the evolution of the overall AGN population \citep{Fanidakis2012}.

This paper explores the predictions of the \galform model for the host DM halo mass of X-ray AGN as a function of $2-10\,\rm{keV}$ accretion luminosity and redshift. Compared to previous versions of \galform, the cosmological parameters have been updated to values similar to those determined by the WMAP7 data \citep{Komatsu2011}. In particular, the rms density fluctuation on scales of $8\,h^{-1}$\,Mpc is set to $\sigma_8=0.8$ compared to the value of $\sigma_8=0.9$ used in earlier \galform models. The model agrees well with observations of galaxies in the local Universe. Its best-fitting parameters in the WMAP7 cosmology, along with the resulting predictions for galaxy LFs and number counts will be presented in a forthcoming publication (Lacey \etal in prep).

\subsection{The growth of BHs}

\galform uses a hybrid BH accretion and galaxy formation model as described in \citet{Fanidakis2012}. In this model the BH growth is coupled to the evolution of its host galaxy and DM halo. The code distinguishes between two modes of black hole fuelling, the starburst mode, which relates to the dynamical and merger history of the host galaxy, and the hot-halo mode, which is associated with the diffuse gas in the DM halo. We briefly summarise below the main characteristics of each mode.

\paragraph*{Starburst mode:} In this mode, the build-up of the BH mass is tightly correlated with the mass of cold gas that turns into stars during a burst of star formation. Starbursts in \galform occur when the host galaxy experiences a disk instability, a major galaxy merger or a minor merger in a gas rich disk. These processes are assumed to involve the entire cold gas reservoir of galaxies in a starburst. Due to the catastrophic impact of those processes on the galaxy morphology it is further assumed that these processes are efficient in driving cold gas towards the inner parts of the galaxy and therefore providing the central BH with fuel. The amount of gas that is accreted by the BH during a starburst is a fraction, \fbh, of the total gas mass that turns into stars. \fbh is constrained by fitting the $\Mbh-\Mbulge$ correlation and BH mass function at $z=0$ (in the model of \citealt{Bower2006}, $\fbh=0.5\%$). The starburst mode is associated with intense and luminous accretion. It is responsible for building the bulk of BH mass in \galform and quasars (considered to be AGN with $\Lbol>10^{46}\ergsec$) are active exclusively during this mode \citep{Fanidakis2012}.

\begin{figure*}
\center
\includegraphics[scale=0.74]{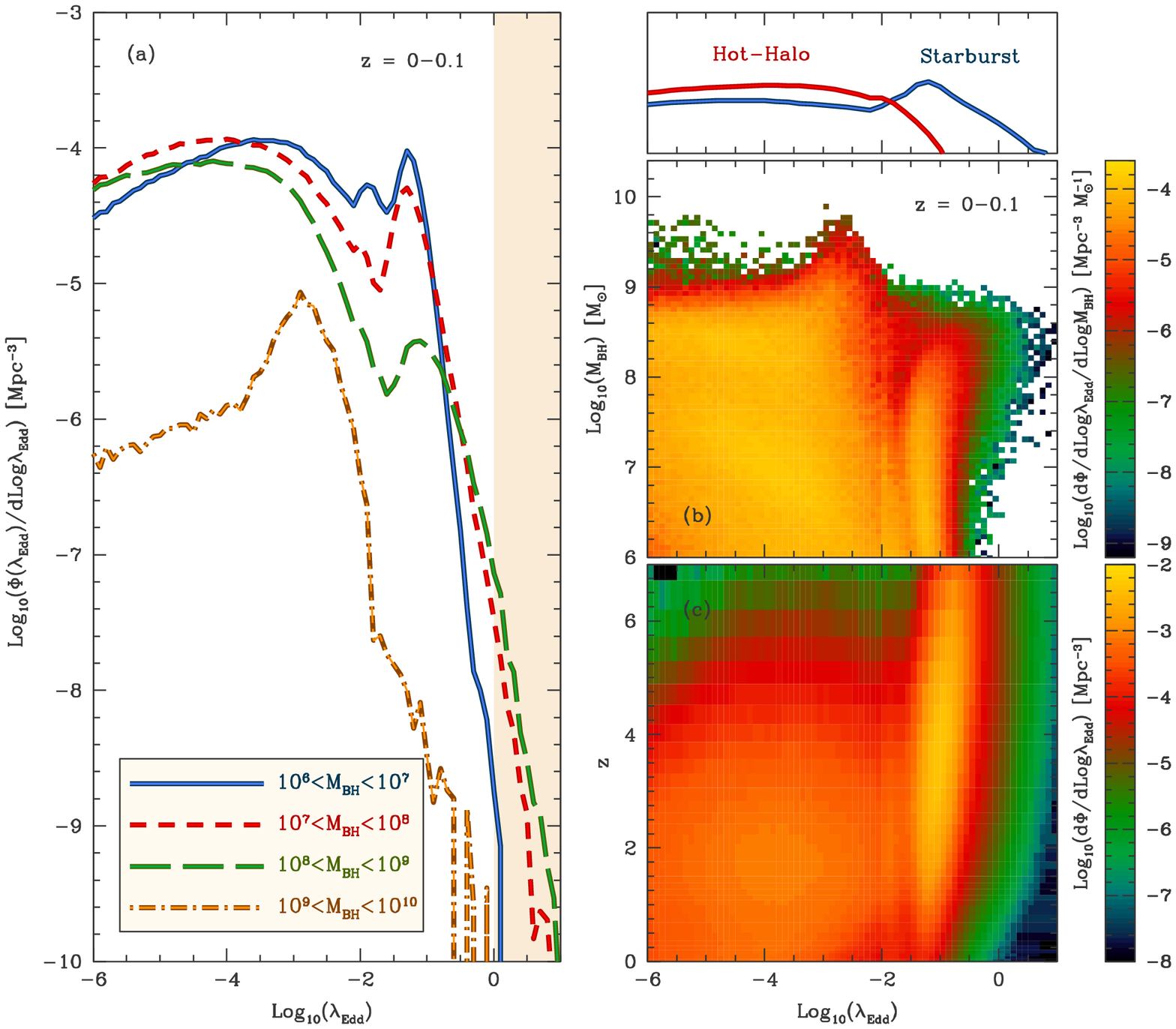}
\caption{(a) The distribution function of $\ledd$ at $z=0-0.1$ in four different BH mass bins, as indicated by the key. The shaded area indicates the super-Eddington regime. (b) The density of accreting BHs (in $\mpc {\rm d}\log\Msun^{-1}$) in the $\lgledd-\log_{10}\Mbh$ plane at $z=0-0.1$. The histograms on top of the panel show the $\ledd$ distribution function for AGN in the hot-halo (red) and starburst (blue) modes. (c) The two-dimensional volume-weighted histogram showing the evolution of the $\lgledd$ distribution as a function of $z$. The different colour shading corresponds to the density of objects in 
 a given $\ledd$ bin, as indicated by the colour bar on the right.} 
\label{ledd_mbh}
\end{figure*}

\paragraph*{Hot-Halo mode:} In this mode, gas is accreted onto the BH directly from the diffuse gas in the DM halo, without first being cooled into the galactic disk. For this to happen, it is necessary that the gas has reached hydrostatic equilibrium within the gravitational potential of the halo and has formed a quasi-static hot atmosphere. In this case, the cooling time of the gas at the cooling radius (the point where the cooling time is equal to the age of the halo) is longer than its free-fall time at this radius. Typical haloes where this condition is satisfied have masses greater than $\sim10^{12.5}\Msun$. In these haloes, the model invokes AGN activity to balance the cooling of gas. As a consequence, the hot-halo accretion mode is coupled to AGN feedback. The heating energy is taken to be a fraction, $\epsilon_{\rm BH}$, of the Eddington luminosity of the BH, $\Ledd=1.4\times10^{38}\Mbh~\ergsec$; if this luminosity exceeds the cooling luminosity, \Lcool, the cooling of gas is suppressed. The gas accreted by the BH during the process of cooling suppression is tuned to the amount needed to produce a luminosity output equal to \Lcool~ (i.e., $\dot{M}_{\rm BH}\sim\Lcool/c^2$, see \citealt{Bower2006}; \citealt{Fanidakis2012}, for further details). The accretion luminosity in this mode becomes important only in very massive haloes ($\Mhalo\gtrsim10^{14}-10^{15}\Msun$), where the cooling luminosity is relatively high (see Section~3.3). We note that \galform is currently the only model that includes a calculation of the AGN luminosity produced during the accretion of gas in the hot-halo mode.

At every timestep, \galform computes the amount of gas accreted during the starburst mode (given that a disk instability or galaxy merger has taken place) and hot-halo mode (if the AGN feedback conditions are satisfied). The gas accreted during the starburst mode is converted into an accretion rate by assuming that the accretion duration is proportional to the dynamical timescale of the host spheroid. In the hot-halo mode the accretion rate is calculated using the timestep over which gas in accreted from the halo. The bolometric luminosity of the accretion flow, $\Lbol$, is then calculated by coupling the accretion rate with the Shakura-Sunyaev thin disk solution \citep{Shakura1973} for accretion rates higher than $1$ percent of the Eddington accretion rate (i.e., the mass accretion rate in Eddington units of   $\dot{m}\geqslant 0.01$) or, otherwise, the ADAF thick disk solution \citep{Narayan1994}. We further refer the reader to \cite{Fanidakis2012} for a detailed account of the accretion physics of the model.

\subsection{The Eddington ratio, $\ledd$}

\begin{figure*}
\center
\includegraphics[scale=0.88]{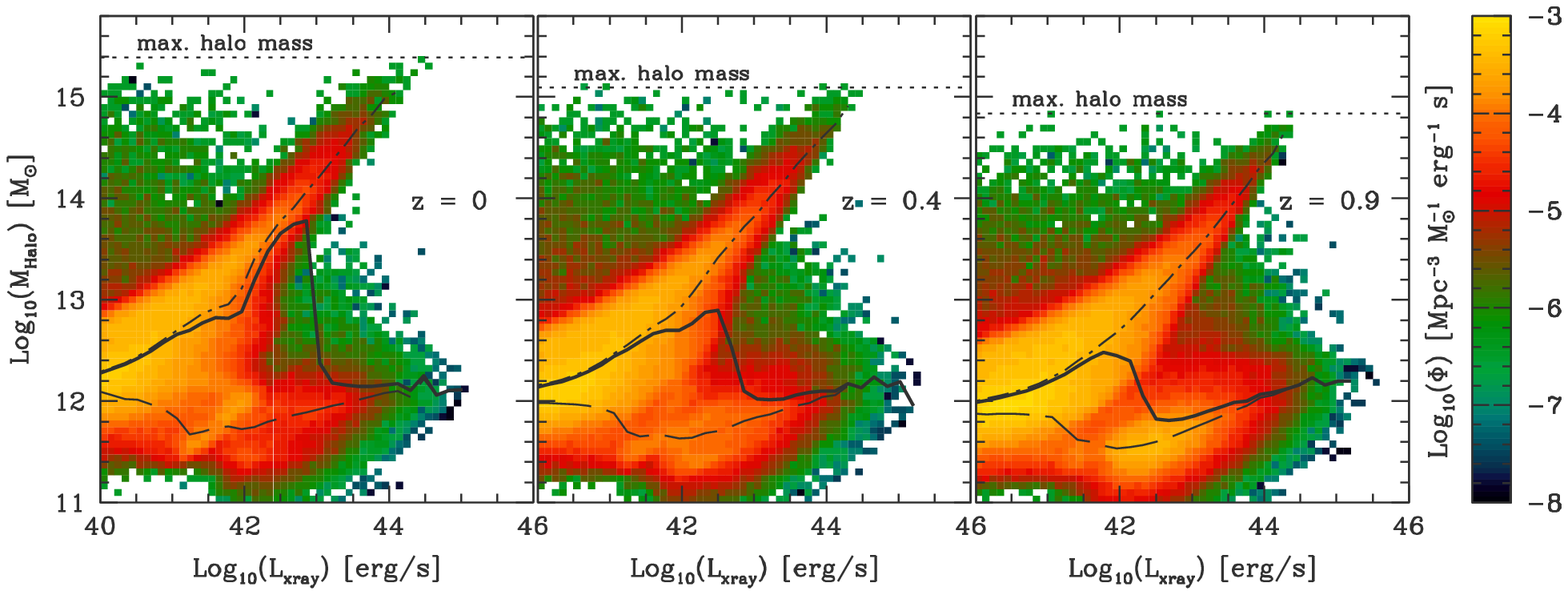}
\caption{The two-dimensional volume weighted histogram of $\Lxray$ ($2-10$~keV) and $\Mhalo$ at $z=0, 0.4$ and $0.9$. The solid line in every panel indicates the median of the $\Mhalo-\Lxray$ correlation. To guide the reader through the locus of each mode, we plot the median halo mass of AGN in the starburst (dashed lines) and hot-halo (dot-dashed lines) mode separately. The dotted line indicates the mass of the most massive halo in place at that redshift.} 
\label{lxray_mhalo}
\end{figure*}

Important insights into the properties of the two accretion modes in \galform are obtained by studying the distribution of the Eddington ratio, $\ledd$, defined as the accretion rate, $\dot{M}_{\rm BH}$, relative to the Eddington value, $\Medd$. $\ledd$ is calculated as in \citet{Fanidakis2012}, taking into account the transition from ADAFs to thin disks at $\dot{m}=0.01$ and the logarithmic dependance of $\Lbol$ on the accretion rate in the super-Eddington regime, $\Lbol\propto\ln(1+\dot{m})\Ledd$. We note that in the regime $0.01<\dot{m}<1$ the value of $\ledd$ is simply $\dot{m}$. For $\dot{m}>1$, $\ledd$ scales as $\ln(1+\dot{m})$.

Fig. \ref{ledd_mbh}a shows the \ledd distribution function at $z=0-0.1$ in four different BH mass bins. The plot shows a bimodal distribution for BHs with masses $<10^9\Msun$, with a broad peak in the ADAF regime ($\lgledd<-2$) and a second peak in the thin-disk regime ($\lgledd>-2$). BHs more massive than $10^{9}\Msun$ are found only in the ADAF regime at $z=0-0.1$.

Interestingly, \citet{Kauffmann2009} find a similar \ledd distribution in a large sample of galaxies in the Sloan Digital Sky Survey \citep[SDSS,][]{York2000}. These authors employed the $L\textrm{[OIII]}/\Mbh$ ratio (with $\Mbh$ estimated using stellar velocity dispersions) as a proxy for \ledd and showed that the suggested
distribution has a bump at $L\textrm{[OIII]}/\Mbh\simeq0.1$\footnote{According to the bolometric corrections assumed by \citeauthor{Kauffmann2009}, the Eddington
limit, i.e., $\log_{10}\ledd=0\,(\equiv L/\Ledd=1)$ corresponds to $\log_{10}(L\textrm{[OIII]}/\Mbh)\simeq1.7$.} for low mass BHs, which is replaced by a power law for more massive BHs. The observations of \citeauthor{Kauffmann2009} span a range of $-1.8\lesssim \log(L\textrm{[OIII]}/\Mbh)\lesssim1.9$, which corresponds to $-3.5\lesssim\log_{10}\ledd\lesssim0.2$ in our plot. A qualitative comparison between their Fig.~5 and our Fig.~\ref{ledd_mbh}a suggests that the complex shape found by \citeauthor{Kauffmann2009} could be a facet of the bimodal nature of \ledd predicted by \galform.

We further explore the bimodality of the \ledd distribution in \galform by plotting in Fig.~\ref{ledd_mbh}b the space density of AGN at $z=0-0.1$ in the two-dimensional $\ledd-\Mbh$ plane. \galform predicts that the bulk of BHs accrete in the ADAF regime ($\lgledd\lesssim-2$). There is only a small fraction of BHs experiencing radiatively efficient accretion, which is represented by the branch around $\lgledd\simeq-1$ extending vertically up along the $\Mbh$ axis. Integrating along the $\Mbh$ axis and distinguishing between accretion in the starburst and hot-halo modes gives the histogram depicted at the top of the $\ledd-\Mbh$ plane. Evidently, the nature of the two modes now becomes clear. The low-\ledd peak is due to the hot-halo mode, while the high log-normal \ledd peak corresponds to the starburst mode. Both modes have a roughly lognormal distribution in $\ledd$, although the starburst mode is also characterised by a long tail extending to very low \ledd values. The convolution between the two modes gives for BH masses below $10^9\Msun$ a bimodal distribution with a strong dip at $\lgledd$, where the two modes intersect.

The relative contribution of each accretion mode to the \ledd distribution function changes with redshift as shown in Fig.~\ref{ledd_mbh}c. AGN in the starburst mode become progressively more abundant with increasing redshift, whereas AGN in the hot-halo mode follow the opposite trend and decrease in abundance. The strong
 evolution with redshift of the starburst mode AGN is a result of  the abundant cold gas supplies present in galaxies at higher redshifts. In  contrast, the abundance of haloes in quasi-hydrostatic equilibrium, and thus susceptible to AGN feedback, which can potentially produce AGN via hot-gas accretion, increases as redshift decreases. 

\begin{figure*}
\center
\includegraphics[scale=0.82]{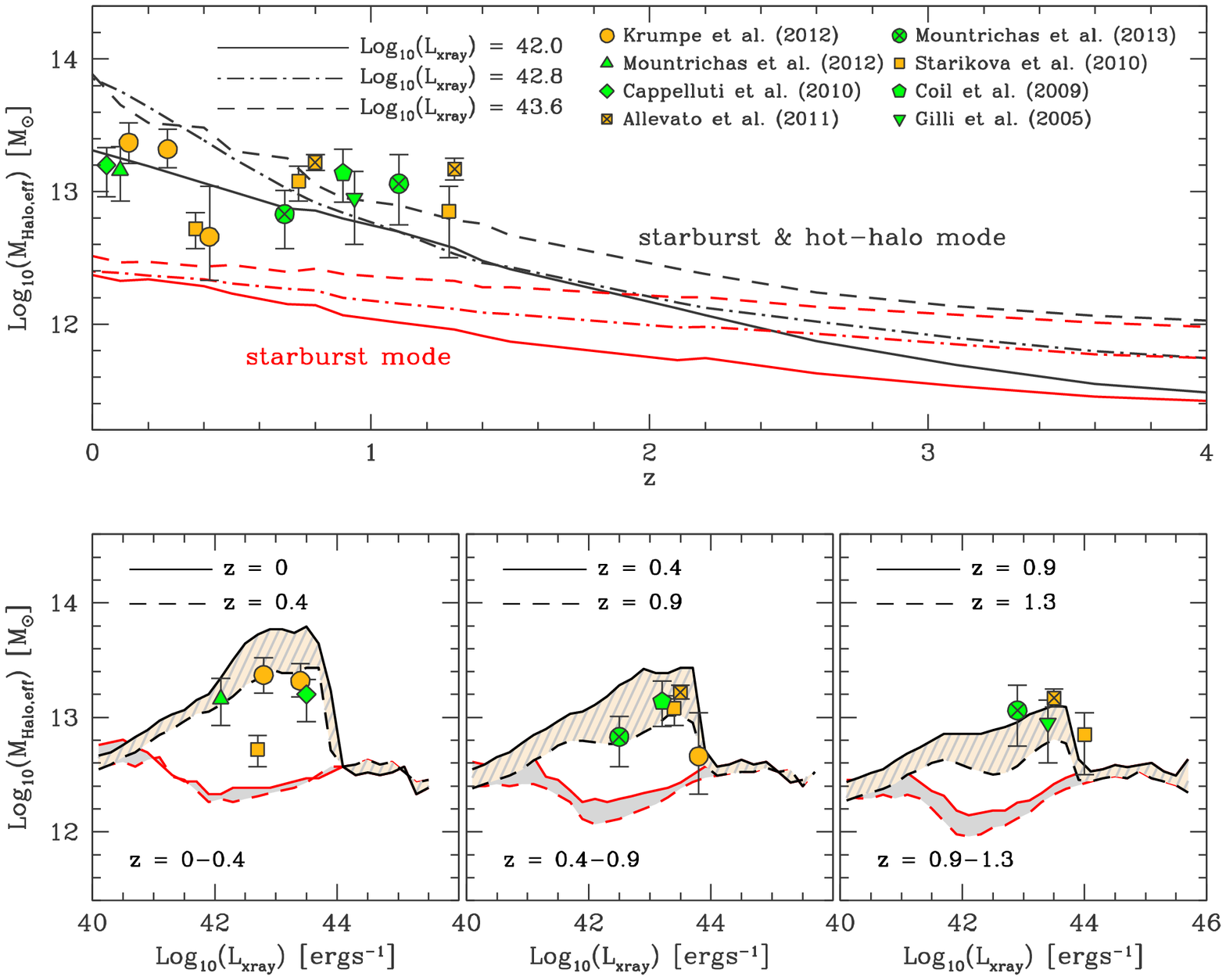}
\caption{The effective host halo mass of AGN, $M_{\rm halo,eff}$, plotted as a function of redshift (top panel) and hard X-ray ($2-10$~keV) luminosity, $\Lxray$ (bottom panels). Top panel: $M_{\rm halo,eff}$ for three different luminosity populations, $\log(\Lxray/(\ergsec))=42,42.8,43.6$, as a function of redshift (black lines). Predictions are compared to the observational estimates of the DM halo mass from Table~1. Values from different X-ray studies are plotted using different symbols, as indicated by the labels. The different colour shadings indicate the X-ray band in which the original measurement was performed: green for the hard ($2-10\,\rm{keV}$) and orange for the soft band ($0.1-2.4\,\rm{keV}$). Also shown are predictions for $M_{\rm halo,eff}$, for AGN accreting only during the starburst mode (red lines). Bottom panels: $M_{\rm halo,eff}$ as a function of hard X-ray luminosity ($2-10$~keV), $\Lxray$. Each panel corresponds to a different redshift interval, as labeled by the key. The top and bottom black lines of the hatched region show $M_{\rm halo,eff}$ as calculated at the lowest and highest $z$ values of the redshift bin. The red lines show the same predictions, but for AGN accreting only during the starburst mode. Observational estimates of the DM halo mass are plotted in the redshift panel that includes the mean redshift of the X-ray sample.}
\label{lxray_mhalo_obs}
\end{figure*}

\subsection{The AGN environment} \label{theoretical_predictions}

The distinct nature of each accretion mode in \galform gives rise to different environmental properties for the starburst and hot-halo AGN population. Because of the link of AGN feedback to the quasi-hydrostatic regime we expect hot-halo AGN to be associated with haloes more massive than $\Mhalo\sim10^{12.5}\Msun$. On the other hand, starburst AGN are characterised by intense accretion involving large amounts of gas. AGN in this mode are found primarily in gas-rich environments ($\Mhalo\lesssim10^{11.5}-10^{12.5}\Msun$), where gas can cool efficiently onto the galactic disk. The brightest AGN (quasars) therefore live in intermediate mass haloes \citep{Fanidakis2013}.

To understand the environmental dependence of the AGN in \galform in more detail, we show in Fig.~\ref{lxray_mhalo} the volume density  of AGN on the two-dimensional plane of DM halo mass and hard X-ray  ($2-10\,\rm{keV}$) luminosity, $\Lxray$, at $z=0, 0.4$ and $0.9$. This quantity is calculated directly from \ledd by applying the bolometric corrections from \citet[][see \citealt{Fanidakis2012} for further details]{Marconi2004}. As illustrated by all the individual redshift panels, AGN have a complex distribution on the $\Lxray-\Mhalo$ plane. Depending on the mode they accrete in, they are either found on the lower-middle part of the plane (starburst mode) or distributed diagonally upwards through the plane (hot-halo mode). In the starburst mode, AGN scatter around haloes of mass $\sim10^{12}\Msun$. Thus, these AGN are associated with average DM environments. The typical progenitor hosts of AGN in this mode are gas-rich disk galaxies that have recently experienced a merger or a disk instability. In contrast, in the hot-halo mode we find a strong (positive) correlation with X-ray luminosity, which extends to halo masses of $\sim10^{15}\Msun$. AGN in this mode typically live in groups, rich clusters and superclusters and are hosted by elliptical galaxies. 

The shape of the two regimes remains the same with increasing redshift, although the relative density of AGN in the two modes changes as expected from Fig.~\ref{ledd_mbh}c. As a consequence, there is a complex dependence of the median host DM halo mass of AGN on accretion luminosity and redshift. For example, at $z=0-0.4$ the median AGN DM halo mass shows a steep increase until $\Lxray\simeq10^{44}\ergsec$, beyond which it drops sharply and flattens to halo masses of $\sim10^{12}\Msun$. However, at higher redshifts the shape of the median changes. The starburst-mode AGN are more dominant in space density and therefore the median remains close to halo masses of $\sim10^{12}\Msun$, the typical halo mass where cold gas accretion dominates.

\section{Comparison with X-ray observations}

The richly varied environmental dependence of AGN in \galform shown in Fig.~\ref{lxray_mhalo} suggests the existence of luminous AGN in a wide range of halo masses. In this section we compare the expected DM halo mass of AGN to the observational estimates from Table~1. To calculate a measure of the host DM halo mass from the model which can be compared with the observational estimates we first compute an effective bias parameter, $\beff$ \citep{Baugh1999}, by weighting the bias parameter $b$ of DM haloes with mass $\Mhalo$ by the mean number of AGN they host, $N_{\rm AGN}$,
\begin{equation}
b_{\rm eff} = \frac{\int b(\Mhalo)N_{\rm AGN}(\Mhalo )n(\Mhalo ){\rm d}\log \Mhalo  }{  \int N_{\rm AGN}(\Mhalo )n(\Mhalo ){\rm d}\log \Mhalo},
\end{equation}
Here $n(\Mhalo )$ is the number density of DM haloes with mass $\Mhalo$.  The bias parameter of a given halo mass, $b(\Mhalo)$, is calculated using the ellipsoidal collapse model of \citet{Sheth2001}. From $\beff$ we then calculate the halo mass, $M_{\rm halo,eff}$, using the effective bias formula, i.e., $b(M_{\rm halo,eff})=b_{\rm eff}$. This method is the same as the one used to infer DM halo masses from the AGN bias in Section~2. Hence, our theoretical predictions and observational estimates for the DM halo mass are consistent with each other. 

We plot $M_{\rm halo,eff}$ (as a function of redshift and $\Lxray$) against the observational constraints from the clustering of X-ray selected AGN (Table~1) in Fig.~\ref{lxray_mhalo_obs}. $M_{\rm halo,eff}$ as a function of redshift (top panel, black lines) is calculated by computing $\beff$ for three different luminosity populations ($10^{42}, 10^{43}$, and $10^{44}\ergsec$, with a bin width of $\delta\log\Lxray=0.1$) and plotting its evolution with $z$. To understand better how the different accretion modes affect the expected host DM mass, we show the same predictions without taking into account the contribution of the hot-halo mode in the calculation of $M_{\rm halo,eff}$. To derive $M_{\rm halo,eff}$ as a function of X-ray luminosity (bottom panels), we derive $\beff (z)$ considering the average number of AGN in a luminosity bin $d\Lxray$. We then compute $M_{\rm halo,eff}$ and plot it as a function of $\Lxray$ in three redshift ranges ($z=0-0.4$, $0.4-0.9$, and $0.9-1.3$), where the redshifts at which $M_{\rm halo,eff}$ is calculated correspond to the boundaries of the redshift range.

In this comparison one should be cautious about a number of potential observational biases and uncertainties in the DM halo mass estimation. Many of the data points in Fig.~\ref{lxray_mhalo_obs} are estimated under the assumption that the auto-correlation function of AGN is a power-law. In reality the clustering signal deviates from this simple functional form. At small scales it is dominated by pairs that belong to the same parent DM halo (1-halo term) and at large scales by pairs in distinct DM haloes (2-halo term). Depending on the relative contribution of the two components and the scale of the pair separation where the 1-halo term becomes dominant, single power-law fits to the clustering signal may overestimate (or underestimate) the typical DM halo mass of AGN (see \citealt{Krumpe2012} for a comparison of the bias parameter as derived from HODs and power-law fits). Additionally, observational determinations of the DM halo mass of AGN that use relatively small area X-ray surveys are often biased high because of sampling variance, i.e., they do not represent the typical (average) Universe.

Finally, each data point in Fig.~\ref{lxray_mhalo_obs} has its own distinct selection function, e.g., redshift range and X-ray flux limit. Therefore, it would have been more appropriate to provide a separate model prediction for each sample in Table~\ref{table:xray_samples}. This approach however, would have made the visualisation of the comparison between model and observations cumbersome. 

Despite these limitations, the agreement between model predictions and observations in Fig.~\ref{lxray_mhalo_obs} is very good. There is some mild tension between the model predictions and certain X-ray samples at $z=0-0.4$ and $z=0.9-1.3$. More data are needed at these redshifts to further investigate whether the \galform model systematically overestimates (or underestimates) the DM halo mass of AGN at these redshifts.

\section{Discussion}

The observational estimates in Fig.~\ref{lxray_mhalo_obs} suggest that moderate X-ray luminosity ($\Lxray\sim10^{42}-10^{44}\ergsec$) AGN inhabit haloes with $\Mhalo\sim10^{13}\Msun$. This is a much higher halo mass than that estimated from observations of UV luminous QSOs in the 2dF and SDSS surveys ($\Mhalo\sim10^{12}\Msun$: \citealt{Croom2005}; \citealt{daAngela2008}; \citealt{Ross2009}; \citealt{Shanks2011}). The environmental difference between the X-ray and UV populations supports multiple modes of BH accretion. Semi-analytic models in which AGN are fuelled by cold gas only via either galaxy mergers or disk instabilities are consistent with the clustering properties of powerful optically selected $z<2$ QSOs \citep[see e.g,][]{Bonoli2009}. However, these models fail to reproduce the clustering and inferred DM halo masses of AGN with moderate X-ray luminosities.  Fig.~\ref{lxray_mhalo_obs} further demonstrates this point by showing the predictions of the \galform model for the expected DM halo mass of X-ray selected AGN in the starburst-mode only (red lines in top and bottom panels). These systems are expected to live in DM haloes in the mass range $10^{12}-10^{12.6}\Msun$ at $z<1.3$, nearly independent of redshift and with only a mild dependence on accretion luminosity. It is the inclusion of the additional hot-gas accretion mode that brings \galform into better agreement with the observed clustering of X-ray AGN. 

At this point we need to stress that the AGN accretion modes undergo strong evolution with redshift (see Fig.~\ref{ledd_mbh}c). The starburst mode becomes the dominant channel at $z\sim3$, while the hot halo mode becomes significantly less important with increasing redshift. From Fig.~\ref{lxray_mhalo_obs}, we find that the environment of moderate luminosity AGN at $z\gtrsim3-4$ is shaped almost entirely by the starburst mode. In this case, the model predicts that lower luminosity AGN reside in haloes of lower mass compared to higher luminosity AGN. This applies also to the brightest AGN ($\Lxray>10^{46}\ergsec$), therefore, we expect the environment of the most luminous quasars to be more massive than that of the moderate luminosity AGN \citep[see][]{Fanidakis2013}.

\begin{figure}
\center
\includegraphics[scale=0.42]{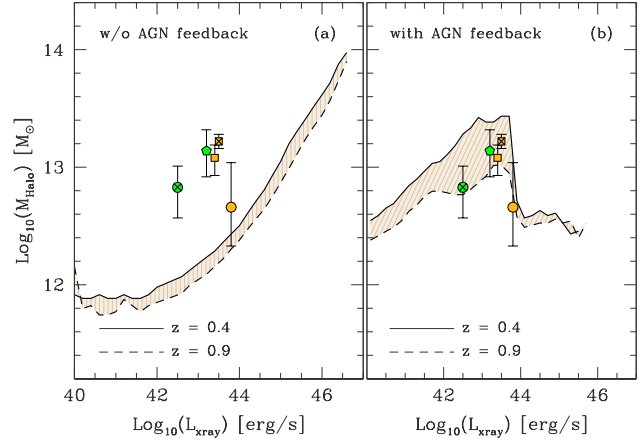}
\caption{The effective host halo mass of AGN, $M_{\rm halo,eff}$, as a function of hard X-ray luminosity ($2-10$~keV), $\Lxray$,  at $z=0.4-0.9$, when the \galform calculation is performed with (right panel) and without (left panel) AGN feedback. The solid lines in each panel indicate the dependance of $\Mhaloeff$ on $\Lxray$ at $z=0.4$ (top lines) and $z=0.9$ (bottom lines), in a similar way as in Fig.~\ref{lxray_mhalo_obs}} 
\label{lxray_mhalo_no_feedback}
\end{figure}

Given that the hot-halo mode is strongly associated with AGN feedback, one might also expect that AGN feedback is crucial in shaping the environmental dependence of AGN. In \texttt{GALFORM},  the low-rate accretion onto the BHs in quasi-hydrostatic halos, and the subsequent feedback of energy, halts the gas overcooling and suppresses the abundance of starburst AGN, which would otherwise increase with halo mass. The suppression of the starburst mode allows the hot-halo AGN to become the dominant AGN population in haloes with masses greater than $10^{11.5}-10^{12}\Msun$ (and luminosities lower than $10^{44}\ergsec$). To illustrate more clearly the effect of AGN feedback on the two different populations of AGN we show in Fig.~\ref{lxray_mhalo_no_feedback}a the $\Mhaloeff-\Lxray$ correlation at $z=0.4-0.9$ when the process of AGN feedback in \texttt{GALFORM} is switched off. In this case, we do not allow the BH accretion energy to be injected in the halo, and thus, cooling in massive haloes ($>10^{12}\Msun$) is not affected by any source of heating. Note that this would not be considered as a viable model as it overpredicts the number of bright galaxies the local Universe. For comparison we show in Fig.~\ref{lxray_mhalo_no_feedback}b the $\Lxray-\Mhalo$ correlation when feedback is on (as in Fig.~\ref{lxray_mhalo_obs}). 

As shown in Fig.~\ref{lxray_mhalo_no_feedback}a, without AGN feedback, the gas in halos cools quickly and fuels starburst episodes, which trigger intense accretion onto the central BH even in the most massive haloes. The model now predicts a monotonic $\Mhaloeff-\Lxray$ correlation, in which $\Mhaloeff$ increases steeply with increasing luminosity and reaches DM halo masses of $\gtrsim10^{14}\Msun$ for the brightest quasars ($\Lxray\simeq10^{46}\ergsec$). Such a monotonically increasing  correlation implies that in a universe without AGN feedback, the X-ray accretion luminosity and therefore the accretion of gas onto the central BH would increase with increasing halo mass.  This is inconsistent with the observed clustering properties of both moderate  luminosity X-ray AGN (Fig.~\ref{lxray_mhalo_no_feedback}) and optically-selected luminous QSOs.

Ignoring AGN feedback also results into a poor fit to the AGN luminosity function in any band. This is because of the overabundance of very bright AGN predicted when cooling is not suppressed in the massive haloes. In principle, the model can be re-tuned to reproduce, within acceptable limits, the observed luminosity function. However, this is achieved only when the average BH accretion timescale is stretched to values greater than the Hubble time. Even in this case, an acceptable fit in one redshift bin does not guarantee the correct evolution throughout the entire redshift range for which observations are available ($0<z<6$).

Finally, in addressing the issue of the dependence of X-ray luminosity on halo mass, we find that the picture emerging from the observational data in Fig.~\ref{lxray_mhalo_obs} is not very clear. The data suggest only a weak dependence at $z<0.9$, which vanishes at higher redshifts. Similarly, recent observational studies suggest that there is possibly a dependance of clustering on AGN luminosity (\citealt{Coil2009}; \citealt{Krumpe2010b}; \citealt{Cappelluti2010}; \citealt{Koutoulidis2013}, see also \citealt{Hutsi2013} for an interesting theoretical account on the problem), although the evidence for this is not very strong (see e.g., \citealt{Yang2006}; \citealt{Gilli2009}; \citealt{Starikova2011}). Nevertheless, the picture emerging from \galform is very clear. Indeed, the rise and fall of halo mass in the moderate and high luminosity regimes is a very distinct prediction of the model. 

The strong correlation between DM halo mass and luminosity in the hot-halo mode is a consequence of the strong dependence of the accretion rate on the cooling properties of the halo. In particular, since the accretion rate is calculated directly from the cooling luminosity, $\Lcool$, and $\Lcool$ increases with halo mass, BHs in more massive haloes are expected to accrete more gas from the hot halo. The dependence of $\Mhaloeff$ on luminosity is very prominent at $z<0.9$ and is apparent in a wide range of luminosities ($\Lxray\simeq10^{40}-10^{44}\ergsec$). At higher redshifts, the dependence becomes milder, mainly due the decrease in the number density of hot-halo AGN. Unfortunately, in this analysis the picture we obtain from the observations is evidently not strong enough to support a luminosity-dependent halo environment. This may imply that a more homogeneous observational sample (with possibly a wider luminosity baseline) is needed for this purpose. To achieve this it is important to standardise the method in the literature (power-law fits, HODs etc.) with which the bias and its uncertainty is calculated from the observations \citep[see discussion in][]{Krumpe2012}. This will provide a more consistent picture of the AGN clustering and will minimise biases related to the assumptions of each method.

\section{Summary and Conclusions}

In this analysis, we have compared the halo masses of moderate luminosity AGN, using samples of X-ray selected AGN from the literature, to the theoretical predictions of the galaxy formation model \texttt{GALFORM}. The typical DM halo  mass is re-estimated for all AGN samples in a uniform manner from the bias parameter, $b$, in order to provide a more direct comparison between theory and different observational surveys. The observations indicate that the average DM halo  mass of moderate luminosity ($\Lxray=10^{42}-10^{44}\ergsec$) AGN at $z\simeq0-1.3$ is $\sim10^{13}\Msun$. The comparison with \galform shows very good agreement with observations. The foundation of this agreement is the incorporation of the AGN feedback mechanism in the \galform model and the two modes of AGN accretion; the starburst mode (cold accretion) and hot-halo mode (hot accretion). This is the first time that a galaxy formation model (in which the formation and evolution of galaxies and BHs is fully coupled) can give a physical explanation to why moderate luminosity X-ray selected AGN show a higher clustering strength than UV luminous quasars.

The AGN feedback prevents gas from cooling in very massive DM haloes ($\gtrsim10^{12.5}\Msun$), establishing the starburst accretion mode (disk instabilities and galaxy mergers) as an inefficient AGN triggering mechanism in such haloes. As a consequence, extremely luminous quasar activity is prohibited in the most massive DM haloes. In this case, an alternative fuelling channel rises to dominance, namely the hot-halo mode. In this mode, BHs accrete hot gas from the surrounding hot halo around the galaxy. The low density of the gas initiates relatively slow BH growth and makes the AGN visible at moderate X-ray luminosities. 

The physical interplay of the two accretion modes gives rise to a distribution of Eddington-ratio parameters, which is in good agreement with those inferred from observations. The relative dominance of the hot-halo mode in the low-$z$ universe ($z\lesssim1$) determines the typical $\sim10^{13}\Msun$ DM halo mass that moderate X-ray luminosity AGN inhabit. In contrast, the brightest quasars, which are associated with disk instabilities and galaxy mergers, inhabit $\sim10^{12}\Msun$ DM haloes, namely those haloes in which the intensity of the starburst mode accretion peaks. Due to the strong cosmic evolution that the hot-halo mode undergoes, this picture changes at $z\sim3-4$, where we find the environment of AGN is solely determined by the starburst mode.

Neglecting the process of AGN feedback or the hot-halo mode in our simulations results in a poor match to the inferred average DM halo mass of moderate luminosity X-ray AGN. In particular, in a universe where feedback in massive haloes is switched off, we find that quasars are typically hosted by DM halos with masses $10^{13}-10^{14}\Msun$, while moderate luminosity AGN are found in $\sim10^{12}\Msun$ haloes. Such an environmental dependence is in contrast with what observations suggest.  

Finally, our model suggests a strong correlation between the expected DM halo mass and X-ray luminosity. This dependence becomes particularly evident at $\Lxray\lesssim10^{44}\ergsec$ in the $z\lesssim 1$ universe and originates from the strong coupling of the accretion rate in the hot-halo mode to the cooling properties of DM matter haloes. Although there are signatures of a luminosity dependent environment in the observational samples of X-ray selected AGN that we have compiled in this analysis, these are very weak. Therefore, we argue that more data are needed in order to provide better constraints on how the environment of AGN correlates with luminosity. 

To conclude, in this analysis we have shown the necessity of AGN feedback and the hot-halo mode as an additional accretion mode to galaxy mergers and disk instabilities, for reproducing the correct clustering properties of X-ray AGN. In a future study we will provide an extensive analysis of the clustering properties of moderate and high luminosity AGN and compare directly to the 2-point correlation function, and bias, as estimated for X-ray AGN and UV luminous quasars in past and current surveys.

\section*{Acknowledgements}
The authors would like to thank Aaron Dutton for valuable comments. The authors acknowledge financial support from the Marie-Curie
Reintegration Grant PERG03-GA-2008-230644. Part of this work was supported by the COST Action MP0905 ``Black Holes in a Violent Universe". The simulations used in this paper were carried out on the Cosmology Machine supercomputer at the Institute for Computational Cosmology, Durham. The Cosmology Machine is part of the DiRAC Facility jointly funded by STFC, the Large Facilities Capital Fund of BIS, and Durham University. This work was supported in part by STFC.

\bibliographystyle{mnras}
\bibliography{fanidakis_2013a}

\end{document}